# Negative permittivity and permeability of gold nanorods metamaterials in UV-Vis region

Zahra jalilian[1](✉), Saeideh Edalati-Boostan[2](✉), and Rostam Moradian[1](✉)

[1] Department of Physics, Faculty of Science, Razi University, Kermanshah, Iran.
[2] Department of Physics and Materials Sciences Center, Philipps-Universität Marburg, Marburg, 35032, Germany.

## ABSTRACT
In this article, we report the growth of gold nanorods on glass substrates and copper nanoparticle thin films by cylindrical direct current magnetron sputtering (CDCMS) at room temperature. The grown gold nanorods have short lengths of <20nm and show negative optical parameters in UV-Vis region. So far negative permittivity and permeability were only shown for complex artificial structures. In a case of simple structures like gold nanorods, the negative optical parameters were only predicted by simulation methods and considering ideal structures and they were not yet reported by experimental groups, who has grown or synthesis gold nanorods by physical or chemical methods. The small size of gold nanorods and thickness of our samples compare to other experimental groups could be the reason of negative permittivity and permeability in our case. Low loss metamaterials with simultaneously negative permittivity and permeability are desired for practical applications in many optical devices such as optical switching, waveguides, modulators, and plasmonic antenna arrays. The optical properties of the grown gold nanorods were defined by ultraviolet-visible (UV-Vis) spectroscopy and their quality was assessed through multi-technique characterization using transmission electron microscopy (TEM), field emission scanning electron microscopy (FE-SEM), X-ray diffraction (XRD), and energy dispersed X-ray (EDX).

## 1 Introduction

Materials with negative permittivity and permeability were first studied by a Russian theorist Victor Veselago in 1968 [1]. He hypothesized that negative refraction can occur if both parameters are negative. So far a natural material that can achieve negative values for permittivity and permeability simultaneously has not been found or discovered.

————————————————
Address correspondence to Zahra Jalilian, jalilian.za@stu.razi.ac.ir; Saeideh Edalati-Boostan, saeideh.edalatiboostan@physik.uni-marburg.de; Rostam Moradian, rmoradian@razi.ac.ir

Therefore, it took 33 years to confirm his prediction by fabricating a metamaterial. It is a material engineered to have a property that is not found in nature [2]. In 2000, Smith et al. [3, 4] fabricated a metamaterial with negative refractive index by combining two structures presented by Pendry in 1996 [5] and 1999 [6]. Metallic nanorods responsible for the negative permittivity and split rings resonators responsible for the negative permeability. Low loss metamaterials with simultaneously negative permittivity and permeability are desired for practical applications in many optical devices such as plasmonic antenna arrays [7], epsilon-near-zero metamaterial at visible wavelengths [8, 9], modulators and switches [10], infrared perfect absorbers, photo detecting, electromagnetic energy conversion [11], and as perfect lens in imaging applications [12-14]. Therefore, they have attracted much attention, among them, plasmonic metal-dielectric metamaterials have investigated extensively [15-17]. Plasmonic metamaterials are typically composed of noble metals in which photonic and electronic properties are linked by coupling photons to conduction electrons of metal, known as surface plasmon [18-23]. Noble metals are of particular interest because of their stability and reproducibility. The localized surface plasmon resonances (LSPRs) of metallic nanostructures increase the interactions between photons and materials, which could be controlled by size, shape, and dielectric permittivity of the environment [24]. Fabrication of metamaterials is generally difficult, however, there are some experimental methods including, template-assisted and self-assembled electrochemical deposition [25, 26], conventional solution-based technique [27], electrochemical plating [28, 29], nanolithography [30], stencil deposition techniques [31], etc. In this work, the Au nanorods were grown on glass substrates by cylindrical direct current magnetron sputtering (CDCMS), which has advantages of high growth rate, low-temperature deposition, higher precision, availability, and safety issue of the used materials. Consequently, the optical properties of Au nanorods were studied by violate-visible (UV-Vis) spectroscopy and the effects of environment on respective properties were investigated by sputtering the nanoparticle Cu thin films between nanorods and glass substrates. They show negative permeability and permittivity in UV-Vis region. Moreover, the effect of sputtering time on size and shape of nanorods were examined. To characterize the crystal structure of the samples and define the length of nanorods, X-Ray diffraction (XRD) and transmission electron microscopy (TEM) were used respectively. However, the surface and elemental studies of samples have been done by field emission scanning electron microscopy (FE-SEM) and energy dispersed X-ray (EDX) spectroscopy.

## 2 Experimental details

### 2.1 Instrumentation

Identification and purity of grown nanostructures were done by XRD in wide-angle X-ray scattering (WAXS) using a STADV powder diffraction (STOE, Germany) with a monochromatic radiation from Cu $\kappa_\alpha$ X-ray source ($\lambda$=1.540598 Å). Scintillation counter (detector) operated at 40 kV and 40 mA at a scanning rate of 0.06 °sec$^{-1}$. Morphology and structural of the samples were examined by FESEM, Mira3 (TESCAN, Czech Republic) at 20-30 kV accelerating voltage and TEM CM120 (Philips, Holland) 120kV accelerating voltage with a LaB$_6$ filament as an electron source. Chemical compositions of thin films analysis were done by EDX spectroscopy. Absorption, transmission, real and imaginary parts of the refractive index at the particular wavelength were recorded with a V630 series of UV-Vis double-beam spectrophotometer (JASCO, Japan). UV-Vis spectra were recorded between 300-1000 nm with scan speed 400 nm/min. All of the measurements were carried out at the room temperature (RT) and the glass substrates were used as a reference baseline for all the absorbance measurement.

### 2.2 Materials and methods

Au and Cu targets were used to grow the samples on the glass substrates by CDCMS. Prior to the depositions, the glass slides (Sail brand, CAT. NO. 7120, 1-1.2 mm THICK) were cleaned by ultrasonic cleaning in acetone-ethanol and ethanol-deionized water for 20 minutes following drying in an oven at 80 °C for 1 hour. Moreover, the targets were cleaned by sputter in the Ar (purity 99.995 %) with the pressure of 10$^{-2}$ torr for 10 min. The substrates were at room temperature (RT) and the distance between targets and substrates were kept at 70 mm during sputtering. The chamber was vacuumed to a basic pressure (BP) about 4.4×10$^{-5}$ torr by both of rotary and diffusion pumps prior to the sputtering. Then the pressure was increased to the ambient pressure with an inert gas. Au nanorods were grown from Au target (99.99 % purity, 80 mm diameter, 0.2 mm thickness) in Ar (purity 99.995 %) with working pressure of 3.8×10$^{-2}$ torr, DC voltage of 400 V, and STs of 90, 150, 210, 270, and 330 seconds for sam1-sam5, respectively. However, Cu nanoparticle thin films were grown from a Cu target (99.99 % purity, 80 mm diameter, 1mm thickness) in N$_2$ (purity 99.995 %) with working pressure of 5.3×10$^{-2}$ torrs, DC voltage of 600 V, and fixed sputtering time (ST) of 325 seconds. In the case of Au nanorods/Cu thin films, Au nanorods were deposited on Cu thin films with considering the fact that for each step the sputtering conditions are as what described above. Therefore, sam1#- sam5# correspond to STs of 90, 150, 210, 270, and 330 seconds. The sputtering conditions of different samples are listed in Table 1.

**Table 1** Sputtering conditions

| ID | Sample | BP [torr] | IG | WP [torr] | V [mV] | I [mA] | ST [s] |
| --- | --- | --- | --- | --- | --- | --- | --- |
| Cu | Cu | 4.4×10$^{-5}$ | N$_2$ | 5.3×10$^{-2}$ | 600 | 1.5 | 325 |
| Sam1 | Au nanorods | 8.9×10$^{-5}$ | Ar | 3.8×10$^{-2}$ | 400 | 7 | 90 |
| Sam2 | Au nanorods | 8.9×10$^{-5}$ | Ar | 3.8×10$^{-2}$ | 400 | 7 | 150 |
| Sam3 | Au nanorods | 8.9×10$^{-5}$ | Ar | 3.8×10$^{-2}$ | 400 | 7 | 210 |
| Sam4 | Au nanorods | 8.9×10$^{-5}$ | Ar | 3.8×10$^{-2}$ | 400 | 7 | 270 |
| Sam5 | Au nanorods | 8.9×10$^{-5}$ | Ar | 3.8×10$^{-2}$ | 400 | 7 | 330 |
| Sam1# | Au nanorods/Cu | 8.9×10$^{-5}$ | Ar | 3.8×10$^{-2}$ | 400 | 7 | 90 |
| Sam2# | Au nanorods/Cu | 8.9×10$^{-5}$ | Ar | 3.8×10$^{-2}$ | 400 | 7 | 150 |
| Sam3# | Au nanorods/Cu | 8.9×10$^{-5}$ | Ar | 3.8×10$^{-2}$ | 400 | 7 | 210 |
| Sam4# | Au nanorods/Cu | 8.9×10$^{-5}$ | Ar | 3.8×10$^{-2}$ | 400 | 7 | 270 |
| Sam5# | Au nanorods/Cu | 8.9×10$^{-5}$ | Ar | 3.8×10$^{-2}$ | 400 | 7 | 330 |

## 3  Results and discussion

### 3.1 Structural Properties

The wide-angle XRD of Au nanorods/Cu thin films, sam1#-sam5#, are illustrated in Fig. 1. A strong diffraction peak located around 38° corresponds to the average d-spacing of 2.36Å, which is assigned to the (111) crystallographic plane of face-centered cubic (FCC) of gold (a=b=c= 4.0680 Å). Whereas, the diffraction peaks from (200) and (220) planes are weak and located at 44° and 64°. By increasing ST, more nanorods growth on the planes leading to increasing the intensity. However, increasing the peaks at 37.52°, 38.30°, 37.54°, 38.01° and 38.08° by increasing Au ST demonstrating that the (111) plane is the preferred crystalline growth direction for Au nanorods [32-36].

The weak fixed peak at 2θ around 36° in all of the wide-angle XRD patterns is corresponding to (310) plane due to the average d-spacing of 2.43Å, which belongs to the orthorhombic (a=8.92 Å, b=4.52 Å, and c=2.83 Å) crystalline structure at Au-Cu interface.

The gold nanocrystal average size and lattice strain were defined by Williamson-Hall Eq. 1 [37]:

$$\beta\cos\theta = 4\,\varepsilon\,\sin\theta + \kappa\lambda/D_{WH} \qquad (1)$$

and density of dislocations was obtained from Eq. 2:

$$\delta = 1/D_{WH} \qquad (2)$$

where β is the FWHM, κ is Scherrer constant, $D_{WH}$ is the crystallite size, λ is the X-ray wavelength, ε is the lattice strain, δ is the density of dislocations, and θ is the Bragg angle. The slope of the plot, βcosθ versus 4sinθ, defines the strain, while its intercept, κλ/$D_{WH}$, gives the average crystallite size. Here average crystallite size of sam1#-sam5# are 0.251, 0.801, 1.842, 3.676 and 0.731 nm, respectively. It indicates that crystallite size increases by increasing ST, except for sam5# due to the aggregation of Au nanorods and consequent morphological perturbations. Moreover, it is expected that dislocation density decreases by increasing of crystallite size. It is consistent with results of 3.98, 1.24, 0.54, 0.27, and 1.36 [nm$^{-1}$] for dislocation density of respective samples, sam1#-sam5#. As there is strain around dislocations, the lattice strain of the sam1#-sam5# are calculated, which are -0.22698, -0.04358, 0.00716, 0.07312, and 0.02147, respectively. It is negative for compressive forces, while positive for tensile forces [38]. Since process performed at the RT, the strain could not be affected by growth temperature. However, coalescing of grains and islands during film growth may cause intrinsic strain. Formation of islands close to the surface, minimize the surface energy while inducing a tensile strain.

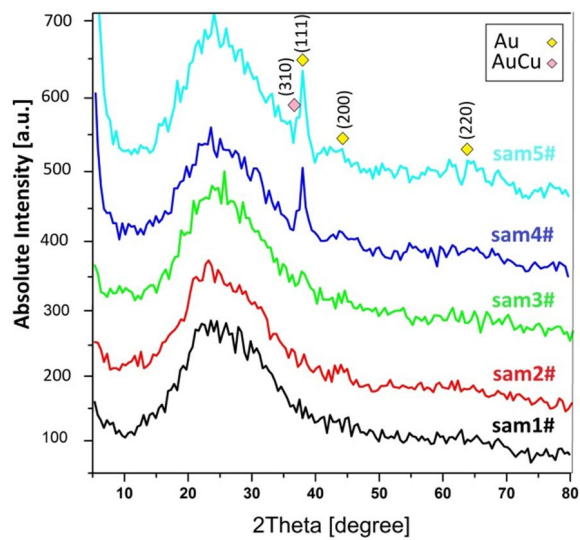

**Figure 1** wide-angle XRD patterns of the Au nanorods/Cu thin films on glass substrates: sam1# (Black), sam2# (Red), sam3# (Green), sam4# (Blue) and sam5# (Cyan).

EDX spectroscopy results of the Au nanorods/Cu thin films for sam1# and sam5# are illustrated in Fig. 2(a)-(e). They clearly demonstrate enhancing of gold intensity by increasing ST. The percentage of Cu and Au is illustrated in the table for each thin film. Small peaks are related to Cu and Au of this films, while larger peaks are related to Ca, Mg, and In of glass substrates. As Si is the main ingredient in glass, the largest peak corresponds to this chemical element.

Figure 3(a)-(e) illustrates FE-SEM top-view images of grown Au nanorods/Cu thin films on glass substrates. It demonstrates the islands of Au nanorods are formed on the Cu nanoparticle thin films and the size of Au nanorods increased as ST increased, which is in agreement with the increment of crystallite size for XRD results. The cross section of FE-SEM image of sam5#, which illustrated in Fig. 3(g), reveals 50nm thickness for Au nanorods. Moreover, the shape and size of the Au nanorods for the same sample were studied by TEM, Fig. 3(f). Gold nanoparticles were found in some spherical and more in rods shapes. According to the scale bar, all the particles have a length less than 20nm.

### 2.1 Optical Response

Figures 4-9 represent the optical response of the Au nanorods-based plasmonic metamaterials.

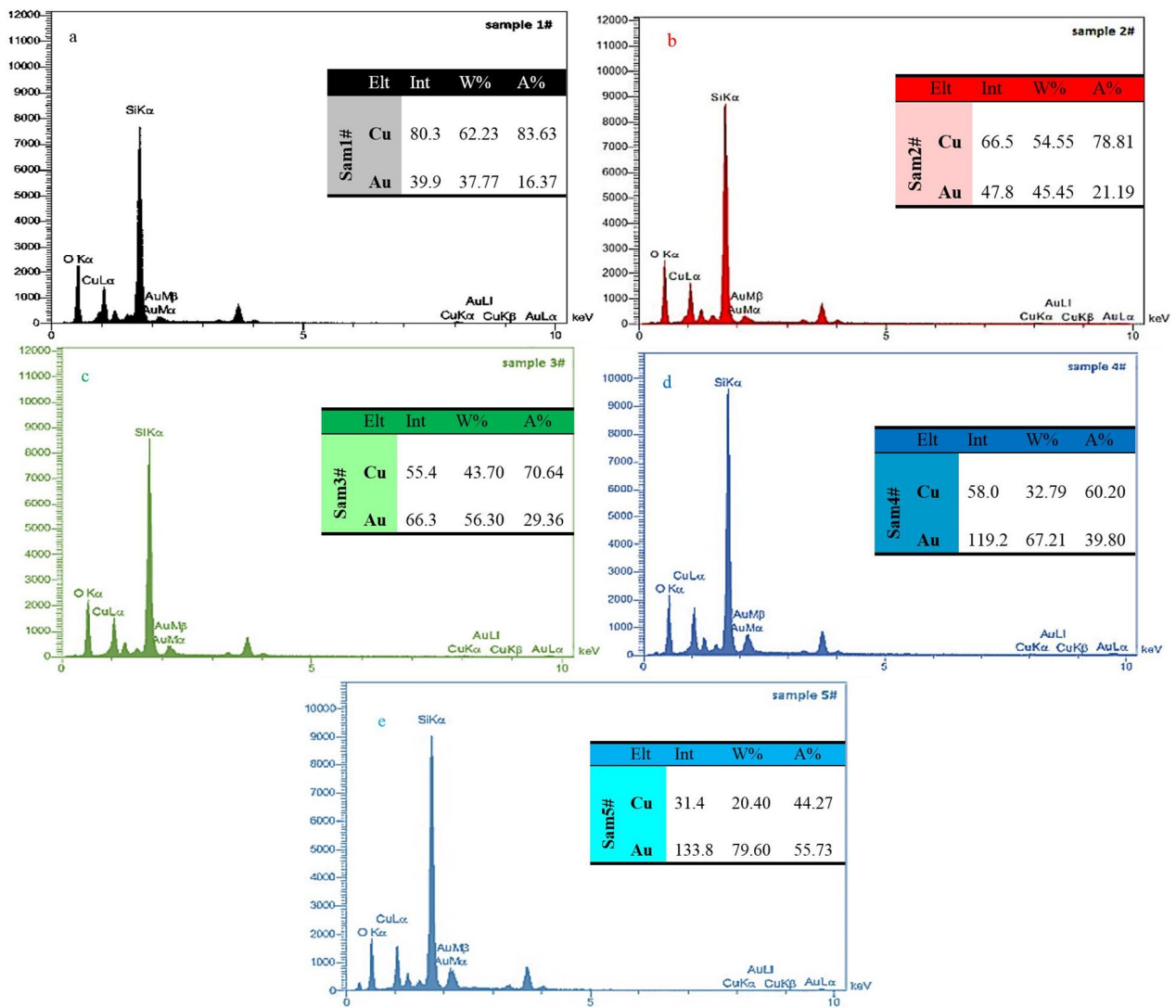

**Figure 2** The EDX spectrum analysis of elements on the surface of Au nanorods/Cu thin films on glass substrates. The atomic and weight percentage values, A% and W%, of Cu and Au are indicated for constant ST of Cu and five different ST of Au nanorods.

For decades, gold nanoparticles have been interesting for scientists due to their ability to tune optical properties by controlling the particle shape, composition, and structure. As predicted by Gan theory in 1915 [39], when the shape of gold NPs change from spheres to rods, the localized surface plasmon resonances band is split into two bands due to the cylindrical symmetry and excitation of collective oscillations of conduction electrons of nanorods. Therefore, instead of one resonance peak due to LSPR, two such peaks observed for Au nanorods. Figure 4(a)-(b) illustrates the variation of absorption spectra of Au nanorods, Cu, and Au nanorods/Cu thin films, which provide information regarding the LSPR. In this work, all samples containing gold nanoparticles exhibit two peaks confirming the growth of nanorods. A weak transverse plasmon resonance (TPR) in the visible region arising from surface plasmon oscillation along the short axis of the rods coinciding with that of nanospheres. Whereas, a strong longitudinal plasmon resonance (LPR) at the higher wavelength is due to the oscillation along the long axis [40-43]. Splitting of plasmon wavelength region could be attributed to the inhomogeneous polarizations parallel and perpendicular to the substrate due to the non-sphericity of the nanorods [44, 45].

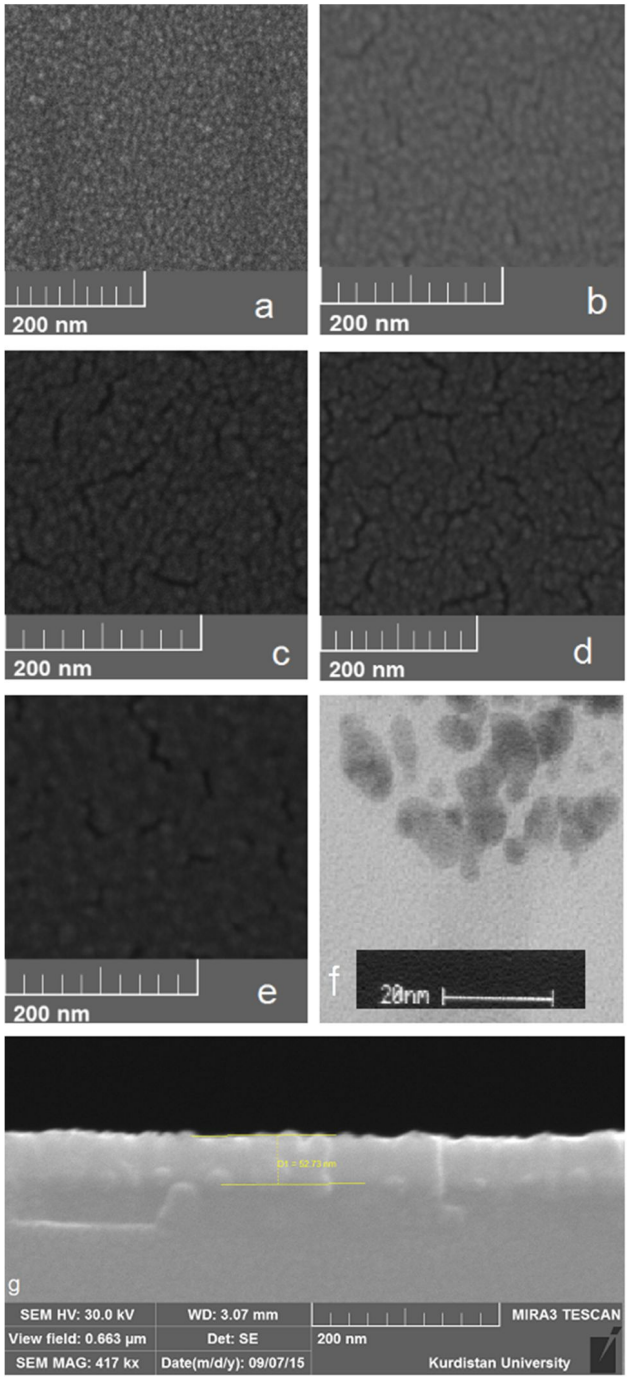

**Figure 3** FESEM images from the Au nanorods/Cu thin films on glass substrates: sam1# (a), sam2# (b), sam3# (c), sam4# (d), sam5# (e) and the cross section view of the Au nanorods on sam5 (g). TEM image of typical Au nanorods in sam5 (f).

TPR is insensitive to the size changes and normally occurs at a fixed wavelength in the visible region. In our case, it is around 420 nm, however, different groups have reported its location varying between 400-500 nm [46, 47]. In contrast, LPR strongly depends on the aspect ratio of the nanorods, which can be understood from Eq. 3. The aspect ratio is the length of the rod divided by its width, R. It has been found that LPR absorption maximum $\lambda_m$ is linearly proportional to the R of the Au nanorods in these thin films [48, 49]:

$$\lambda_m = 54R + 539 \qquad (3)$$

The inhomogeneous polarization enhances higher-order mode oscillations at lower energies [50, 51]. Therefore, LPR is red shifted largely from the visible to the near-infrared region by increasing aspect ratio. These results are in good agreement with previous studies on Au nanorods, where TPR is fixed but LPR occurs at a higher wavelength and redshifts by increasing ST [52-55].

The calculated aspect ratios of Au nanorods thin films are illustrated in Table 2, ranging from 2.83 to 6.68. Furthermore, Au nanorods-based metamaterials exhibit Drude absorption in the ultraviolet (UV) region. As the Au nanorods ST increases, the Drude absorption rises. It could be attributed to the increment of Au nanorods size and consequent electronic overlap between them.

Figure 5(a)-(b) demonstrates transmission spectra of Au nanorods, Cu, and Au nanorods/Cu thin films on glass substrates. All samples containing Au nanorods show two peaks in the range of UV and visible at around 350 and 520 nm, consistent with dips of absorption spectra in Fig. 4 (a)-(b). The peak positions only depend on the type of materials and insensitive to ST or thickness of samples. The absorption edge at 300 nm is due to absorption of glass substrates.

**Table 2** Aspect ratio of the Au nanorods corresponding $\lambda_m$ for samples.

| ID | $\lambda_m$ [nm] | Aspect ratio (R) (Cal. Eq. 3) |
|---|---|---|
| Sam1# | 692 | 2.83 |
| Sam2# | 725 | 3.44 |
| Sam3# | 890 | 6.38 |
| Sam4# | 942 | 7.46 |
| Sam5# | 900 | 6.68 |

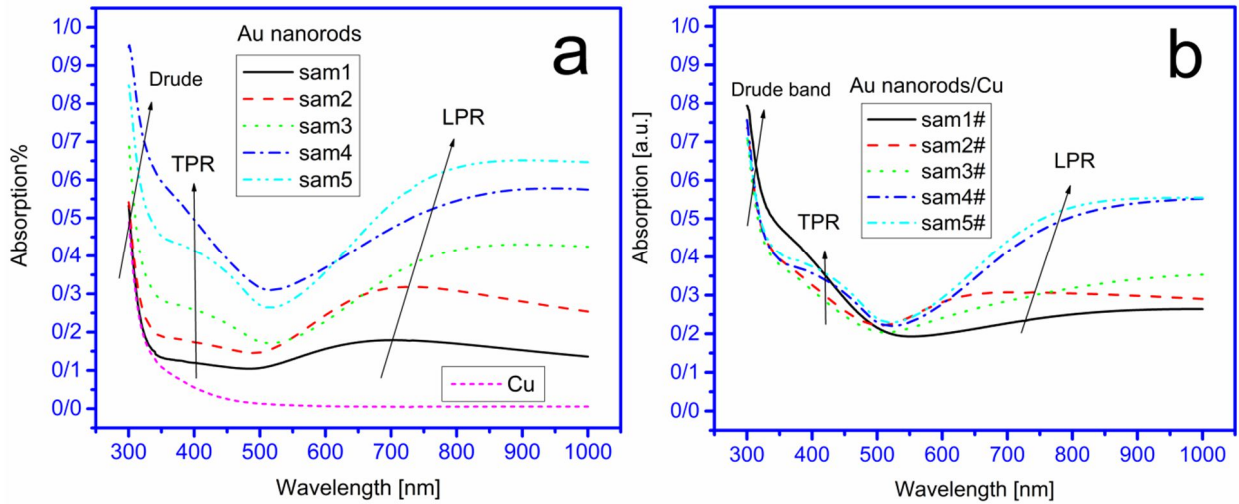

**Figure 4** UV-Vis absorptions spectra of (a) Cu and Au nanorods, (b) Au nanorods/Cu thin films on glass substrates.

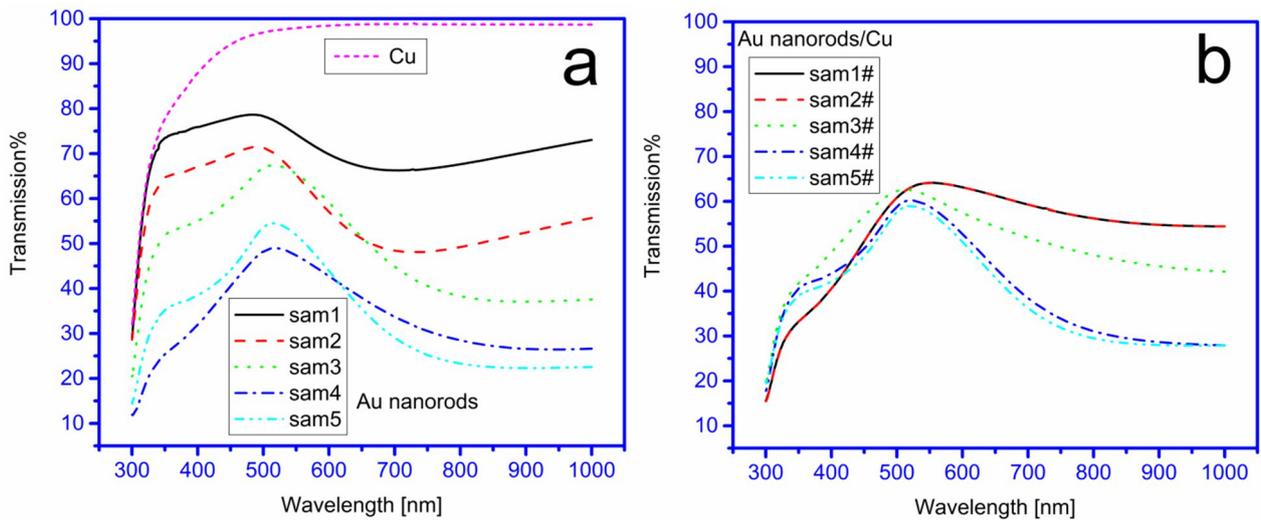

**Figure 5** UV-Vis transmissions spectra of (a) Cu and Au nanorods, (b) Au nanorods/Cu thin films on glass substrates.

Besides the absorption and the transmission spectra of the samples, it is worth exploring their real and imaginary part of refractive index as it is the key parameter in the interaction of light with matter. Figure 6(a)-(b) represents the real part of the refractive index, n, of Au nanorods, Cu, and Au nanorods/Cu thin films. For samples with Au nanorods, n is small at Drude and LSPRs wavelength regions correspond with high absorption at the plasmonic peaks and have a maximum in a visible region around 520 nm. The n of gold nanorod thin films decreases with the increase of ST, except for sam1. These results are consistent with previous reports about the degradation of n with increasing thickness of the thin films [56, 57]. However, in Au nanorods/Cu thin films opposite trend was observed according to the very low refractive i
ndex of Cu. Increasing ST leading to increasing the thickness of gold thin films, which reduces the effect of Cu on the total refractive index of the samples. Moreover, the refractive index dispersion of Au nanorods in presence of Cu becomes weaker. Therefore, such plasmonic metamaterials can act as low or tunable refractive index materials at specific optical wavelengths. While n represents the real part of the complex refractive index, k represents the imaginary part and known as extinction coefficient. When it is positive, the optical depth is positive and diminishes incoming light. However, the negative value of k amplifies incoming light. The extinction coefficient of Au nanorods, Cu, and Au nanorods/Cu thin films are illustrated in Fig. 7(a)-(b). It shows that the spectrum of k, like n, is wavelength dependence and reduces by increasing ST for both Au nanorods, and Au nanorods/Cu thin films. However, linear variation of extinction coefficient with respect to wavelength for pure Cu thin films is found to be non-linear in the presence of Au nanorods. Considering an incident light beam scatters by a collection of electrons, the linear polarization of the medium at a position depend on the applied electric field. The electric field amplitude has been modified by the factor $e^{-\omega kz/c}$, and hence the transmitted intensity will increase with distance in the medium for negative k and decrease for a positive one [58]. It is consistent with our results in Fig. 4, Fig. 5, and Fig. 7, as for the negative k in the range of visible to near-infrared wavelength high transmission and less absorption was observed. There is a region of high tunability in these wavelength ranges, where imaginary part (k) is near zero. This region would be useful for optical switching applications.

The complex refractive index of a material, $N(\omega)=n+ik$ at frequency of $\omega$ is directly related to its relative complex electric permittivity, $\varepsilon(\omega)=(n^2-k^2)+i(2nk)$, and its relative complex magnetic permeability, $\mu(\omega)= \mu_r+i\mu_i$, by Eq. 4 [59-61]:

$$N(\omega)= \pm( \varepsilon(\omega) \mu(\omega))^{1/2} \qquad (4)$$

Therefore, studying variations of permittivity and permeability of gold nanorod-based plasmonic metamaterials with respect to wavelength would provide valuable information in this regard.

Figure 8(a)-(b), and Fig. 9(a)-(b) illustrate the real part of the permittivity and permeability of Au nanorods, and Au nanorods/Cu thin films. Both figures show negative values in the range of UV-Vis for samples with lower ST. The negative value of the real part of relative permittivity could be understood from the Eq. 5:

$$\varepsilon_r = 1 - \omega_{plasma} / \omega \qquad (5)$$

when the frequency of incident light is much lower than the plasmon frequency, $\omega_{plasma} / \omega$ ratio is greater than 1, leading to a negative value. This negative value in metal thin films demonstrates high density of free electrons leading to opaque thin films, while they are transparent for a positive value.

Generally, simultaneous negativity of µ and ε leading to a negative value of n. Although both parameters were negative within the range of UV-Vis in this work, a negative value of n was not observed so far. This could be understood from the necessary condition for n to be negative [62]:

$$\varepsilon_r\mu_i + \varepsilon_i\mu_r < 0 \qquad (6)$$

In a case that both $\varepsilon_r$ and $\mu_r$ are negative, their imaginary counterparts need to be small and positive to full fill this condition. In this work, large negative values of imaginary part of permittivity and permeability prevent observation of negative refraction for the samples. A. N. Grigorenko *et. al.* [63] also reported positive n for their samples with negative values of $\varepsilon_r$ and $\mu_r$ due to the large positive imaginary part of permeability.

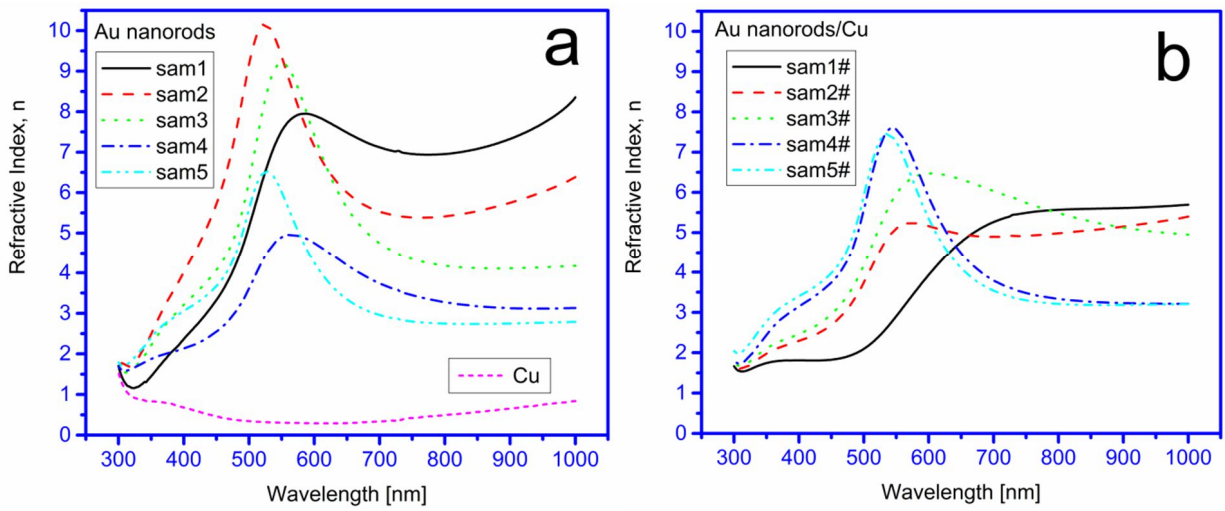

**Figure 6** UV-Vis spectra of the real part of refractive index of (a) Cu nanoparticles and Au nanorods, (b) Au nanorods/Cu thin films on glass substrates.

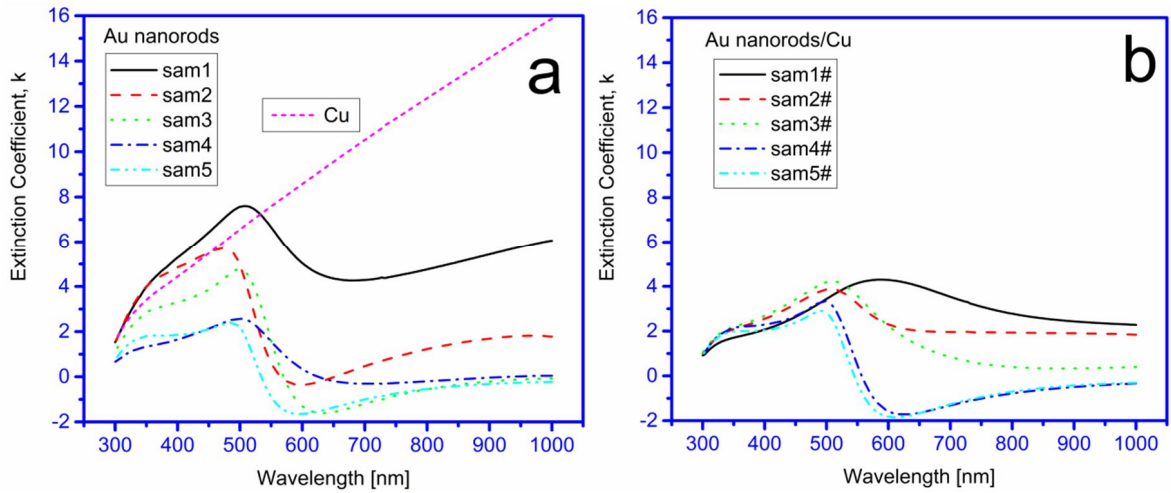

**Figure 7** UV-Vis spectra of the extinction coefficient of the (a) Cu nanoparticles and Au nanorods, (b) Au nanorods/Cu thin films on glass substrates.

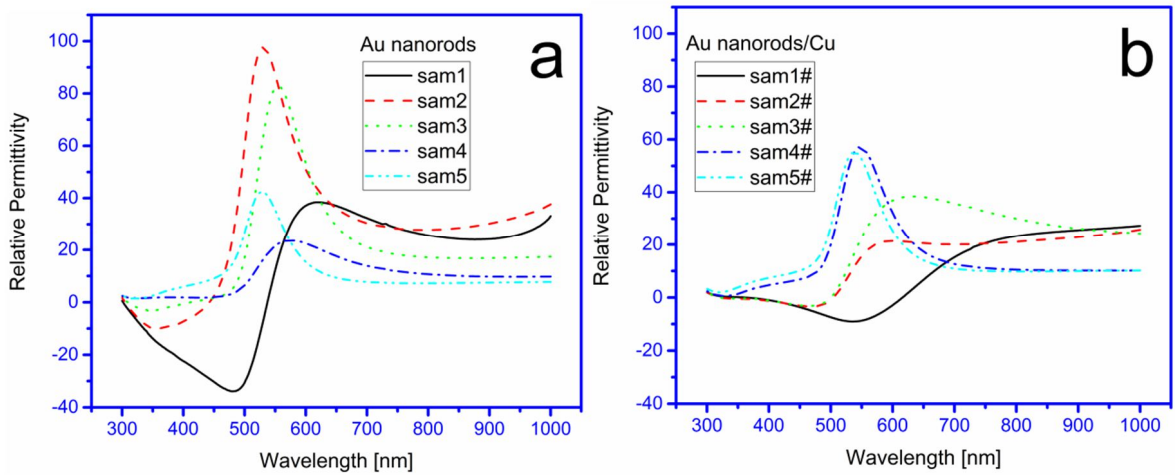

**Figure 8** The relative permittivity of (a) Au nanorods and (b) Au nanorods/Cu thin films on glass substrates.

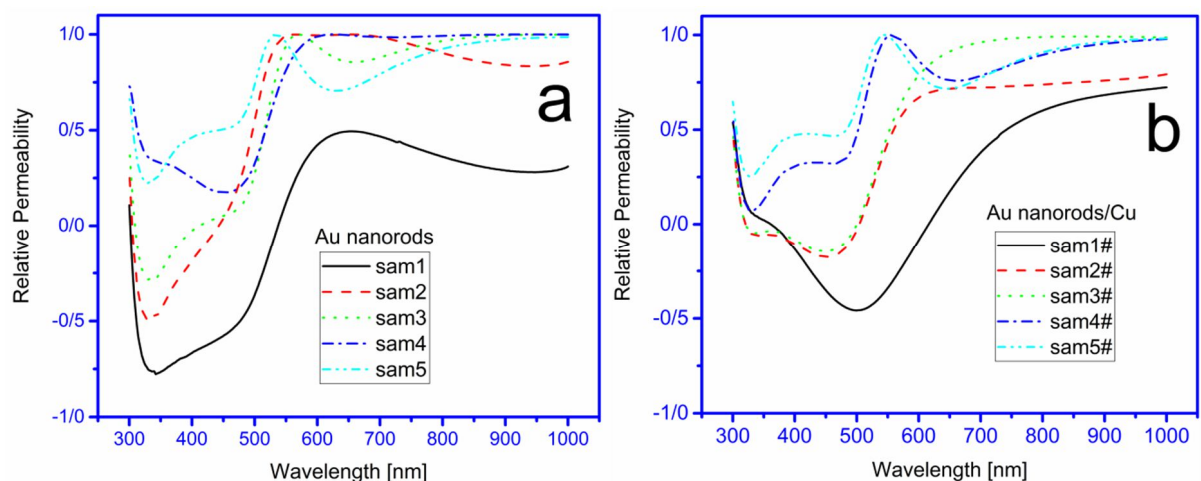

**Figure 9** The relative permeability of (a) Au nanorods and (b) Au nanorods/Cu thin films on glass substrates.

## 3  Conclusion

In this work, we have grown gold nanorods on glass substrates and copper nanoparticle thin films by CDCMS at room temperature. The quality of the samples was determined by TEM, FE-SEM, XRD, and EDX and their optical properties defined by UV-Vis spectroscopy.

Low loss metamaterials with simultaneously negative permittivity and permeability are not found in nature and the artificial structures are generally very complex. In this work, we present simple structures containing gold nanorods, which demonstrate a negative optical response in UV-Vis region. Low loss metamaterials with simultaneously negative permittivity and permeability are desired for practical applications in many optical devices such as optical switching, waveguides, modulators, and plasmonic antenna arrays.

## Acknowledgments

This work was supported by the Iran Nanotechnology Initiative Council number 93700.